# Performance versus Complexity Per Iteration for Low-Density Parity-Check Codes: An Information-Theoretic Approach

Igal Sason, Gil Wiechman

Technion - Israel Institute of Technology, Haifa 32000, Israel. E-mails: {sason@ee, igillw@tx}.technion.ac.il

## Abstract

The paper is focused on the tradeoff between performance and decoding complexity per iteration for LDPC codes in terms of their gap (in rate) to capacity. The study of this tradeoff is done via information-theoretic bounds which also enable to get an indication on the sub-optimality of message-passing iterative decoding algorithms (as compared to optimal ML decoding). The bounds are generalized for parallel channels, and are applied to ensembles of punctured LDPC codes where both intentional and random puncturing are addressed. This work suggests an improvement in the tightness of some information-theoretic bounds which were previously derived by Burshtein et al. and by Sason and Urbanke.

## 1 Introduction

Error-correcting codes which employ iterative decoding algorithms are widely considered state of the art in channel coding. Their outstanding performance and feasible encoding and decoding complexities motivate a theoretical study of the tradeoff between their performance and complexity. While it would be very interesting to explore this tradeoff for finite block lengths, this central issue is still beyond the scope of rigorous analysis. Via information-theoretic bounds, we study this tradeoff in the asymptotic case where the block length tends to infinity. This study is focused on low-density parity-check (LDPC) codes, but we also address other constructions of codes on graphs which closely approach channel capacity with moderate complexity.

LDPC codes are efficiently encoded and decoded due to the sparseness of their parity-check matrices ([18], [19], [20]). An ensemble of LDPC codes is defined by a pair of degree distributions $\lambda(x) = \sum_{i \geq 2} \lambda_i x^{i-1}$ and $\rho(x) = \sum_{i \geq 2} \rho_i x^{i-1}$ where $\lambda_i$ ($\rho_i$) is the fraction of edges in the bipartite graph adjacent to variable (parity-check) nodes of degree $i$. Consider the number of ones in a parity-check matrix which represents a binary linear block code, and normalize it per information bit (i.e., with respect to the dimension of the code). This quantity (defined as the *parity-check density* [21]) is linearly proportional to the number of messages produced in each iteration of the message-passing iterative (MPI) decoder, normalized per information bit. Therefore, the density of the parity-check matrix can be regarded as a measure for the decoding complexity per iteration of LDPC codes normalized per information bit. Our study of the tradeoff between performance and complexity per iteration is initiated by two questions which were addressed by Burshtein et al. [2] and Sason and Urbanke [21]:

*Question 1: How sparse can parity-check matrices of binary linear codes be, as a function of the gap between the code rate for which reliable communication can be achieved and the channel capacity?*

*Question 2: How good can LDPC codes be (even under maximum-likelihood (ML) decoding), as a function of their degree distributions?*

The answer to the first question was given in [21] in the form of lower bounds on the parity-check density (which serve as lower bounds on the decoding complexity per iteration); these lower bounds were later improved by the authors [26], and the improved bounds will be introduced in this paper. Since the density of a parity-check matrix is related to the number of fundamental cycles in the corresponding bipartite graph (see [21]), this also gives a quantitative measure to the statement that bipartite graphs representing good error-correcting codes should have cycles [3]. Referring to the second question, the answer is given in the form of upper bounds on the achievable rates whose derivation is based on performance bounds under ML decoding. Comparing these upper bounds with exact thresholds under iterative decoding, calculated using density evolution (DE) [19], leads to a quantitative measure of the inherent loss in performance due to the sub-optimality of MPI decoding. The derivation of these bounds also allows to derive lower bounds on the bit/block-error probability under ML decoding (see [21], [26]).

In his thesis [4, Theorem 3.3], Gallager provided an upper bound on the rate of right-regular LDPC codes (i.e., LDPC codes with a constant degree ($a_R$) of the parity-check nodes) which is a necessary condition for achieving reliable communications over a binary symmetric channel (BSC), even under optimal ML decoding. This bound shows that right-regular LDPC codes with fixed right degree cannot achieve the channel capacity on a BSC, even under optimal ML decoding. The inherent gap between the achievable rates and the channel capacity is well approximated by an expression which decreases to zero exponentially fast in $a_R$. Burshtein *et al.* have generalized Gallager's bound for general ensembles of LDPC codes transmitted over memoryless binary-input output-symmetric (MBIOS) channels [2]; to this end, they applied a two-level quantization to the log-likelihood ratio (LLR) of these channels which essentially turns them into a BSC.

In [8], Khandekar and McEliece have suggested to study the encoding and decoding complexities of ensembles of iteratively decoded codes on graphs as a function of their gap to capacity. They conjectured that if the achievable rate under MPI decoding is a fraction $1-\varepsilon$ of the channel capacity, then for a wide class of channels, the number of iterations required for decoding with vanishing bit error probability scales like $\frac{1}{\varepsilon}$ and the density of the parity-check matrix scales like $\ln\frac{1}{\varepsilon}$. We note that recent findings on the matching condition for GEXIT curves of capacity-approaching codes seem to suggest an analytical tool for proving the conjecture that the average number of iterations scales like the inverse of the multiplicative gap ($\varepsilon$) to capacity [12]. Based on this conjecture, it follows that for irregular repeat-accumulate codes, the encoding complexity scales like $\ln\frac{1}{\varepsilon}$ and the decoding complexity scales like $\frac{1}{\varepsilon}\ln\frac{1}{\varepsilon}$. The only exception is the binary erasure channel (BEC) where the decoding complexity behaves like $\ln\frac{1}{\varepsilon}$ (same as encoding complexity) because of the absolute reliability of the messages passed through the edges of the graph (hence, every edge can be used only once during the iterative decoding process).

Based on the work in [2], Sason and Urbanke derived an information-theoretic lower bound on the asymptotic density of parity-check matrices [21, Theorem 2.1] where this bound applies to every MBIOS channel and every sequence of binary linear block codes achieving a fraction $1-\varepsilon$ of the channel capacity with vanishing bit error probability. It holds for an arbitrary representation of full-rank parity-check matrices for these codes, and is of the form $\frac{K_1+K_2 \ln\frac{1}{\varepsilon}}{1-\varepsilon}$ where $K_1$ and $K_2$ are constants which only depend on the channel. For the BEC they used considerations relating to the rank of the parity-check matrix to get an improved bound, which is also logarithmic in $\frac{1}{\varepsilon}$. A natural question arises:

*Question 3: Is the logarithmic behavior of the lower bound on the parity-check density true, or is it just an artifact of the bounding technique?*

Sason and Urbanke showed that for any MBIOS channel, the logarithmic behavior of the parity-check density is achievable under ML decoding [21, Theorem 2.2]. A recent analysis in this direction was also presented in [6]. For the BEC, they showed that the sequence of right-regular ensembles of Shokrollahi [24] achieves the improved bound (up to a small additive constant) even under (sub-optimal) MPI decoding [21, Theorem 2.3].

In [10], Measson *et al.* derived an upper bound on the thresholds of ensembles of LDPC codes under maximum a posteriori (MAP) decoding where the transmission of these codes takes place over the BEC. Their general approach relies on extrinsic information transfer (EXIT) charts. Generalized extrinsic information transfer (GEXIT) charts were recently introduced by Measson *et al.* ([9], [12]); they form an extension of EXIT charts to general MBIOS channels, and they fulfill the area theorem. This preservation of the area theorem for general MBIOS channels enables to get upper bounds on the thresholds of turbo-like ensembles under MAP decoding. Under some conditions, the bound in [10] was shown to be tight for the BEC.

A new method for analyzing LDPC codes and low-density generator-matrix (LDGM) codes under bit-MAP decoding is introduced by Montanari in [11]. The method is based on a rigorous approach to spin glasses, and allows a derivation of an upper bound on the achievable rates of these codes under bit-MAP decoding. The computational complexity of this bound grows exponentially with the maximal right and left degrees (see [11, Eqs. (6.2) and (6.3)]). Since the bounds in [10], [11] are derived for ensembles of codes, they are probabilistic in their nature; based on concentration arguments, they hold asymptotically in probability 1 as the block length goes to infinity. It was conjectured that the bounds in [11], which hold for general LDPC and LDGM ensembles over MBIOS channels, are tight.

In [1] and [25], Ardakani *et al.* suggest an approximated one-dimensional analysis which enables to optimize (numerically) the degree distributions for achieving a good tradeoff between performance and complexity.

The derivation of the results presented in this paper was motivated by the fact that the information-theoretic bounds in [2] and [21] rely on a two-level quantization of the LLR and do not take into consideration the entire statistics of the communications channel. In this paper we present a new information-theoretic upper bound on the achievable rates of LDPC codes transmitted over MBIOS channels. This bound holds for optimal ML decoding and is therefore valid under any sub-optimal decoding algorithm. We also discuss an improvement of the information-theoretic lower bound on the parity-check density. These bounds hold for every sequence of codes and do not rely on probabilistic arguments. The bounds are generalized for the case where the codes are transmitted over a set of independent parallel MBIOS channels, where each code bit is a-priory assigned to one of the parallel channels. This setup can be used to model different scenarios which include non-uniform error protection [16], punctured LDPC codes [5], [17] and LDPC-coded modulation schemes. The bounds for parallel channels are used to bound the performance and the decoding complexity per iteration of punctured

LDPC codes transmitted over a single MBIOS channel.

Due to space limitations, the proofs of the statements in this paper are relegated to the full paper versions (see [23] and [26]); further discussion on the relations between these theorems and some previously reported results from [2], [14], [16], [21] are presented there.

## 2 Results for an MBIOS Channel

In this section we present new upper bounds on the achievable rates of LDPC codes, and new lower bounds on their bit-error probability. The transmission of the codes is assumed to take place over an MBIOS channel and the bounds refer to ML decoding. We also consider an improved lower bound on the parity-check density of a sequence of LDPC codes which achieve a fraction $1-\varepsilon$ of the capacity of an MBIOS channel with vanishing bit error probability. These bounds tighten the results in [2], [21]. The upper bound on the achievable rates and the lower bound on the parity-check density in [2], [21] are both derived from a lower bound of the conditional entropy of the transmitted codeword given the received sequence at the output of the communication channel. For the derivation of this bound Burshtein *et al.* considered the entropy of the syndrome of the received sequence, this requires that the received sequence be binary. Hence, a two level quantization which essentially turns the MBIOS channel into the BSC was applied to the LLR at the output of the communication channel. This quantization causes an inherent loss in the tightness of the bounds in [2], [21].

We introduce two approaches to circumvent the two-level quantization of the LLR. The first approach forms a generalization of the approach introduced by Bursthein *et al.* and allows a finer quantization of the LLR to a number of levels which is an arbitrary integer power of 2. Each quantization level is mapped to a member of the Galois field $GF(2^d)$ (where the number of quantization levels is $2^d$ and $d \in \mathbb{N}$), and the concept of syndrome is generalized for members of $GF(2^d)$. These bounds are subject to optimization of the quantization levels, as to achieve the tightest bounds within their form. For optimally chosen quantization levels it can be shown that the bounds become tighter as the number of levels grows. A second approach is to consider the soft value of the LLR which forms a sufficient statistics of ML decoding. In this approach the LLR is split into its sign and its absolute value, due to the symmetry of the channel it is possible to define the syndrome w.r.t the sign of the LLR and use its absolute value as side information. This second approach leads to bounds which are uniformly tighter than the first approach and require no optimization. However, comparing the results of both approaches gives insight on the effect of quantization of the LLR at the decoder due to implementation restrictions. Due to space limitations we only present bounds which are based on the second approach. Some numerical results, presented at the end of this section, demonstrate the results given by both approaches.

As noted above, the bounds on the achievable rates, the bit error probability and the parity-check density are derived from a lower bound on the entropy of the transmitted codeword conditioned on the received sequence. We present an improved lower bound on this entropy, as compared to the one derived in [2].

*Theorem 2.1:* Let $\mathcal{C}$ be a binary linear code of length $n$ and rate $R$ transmitted over an MBIOS channel. Let $\mathbf{x} = (x_1, x_2, \ldots, x_n)$ and $\mathbf{y} = (y_1, y_2, \ldots, y_n)$ be the transmitted codeword and the received sequence, respectively. For an arbitrary representation of the code $\mathcal{C}$ by a full-rank parity-check matrix, let $\Gamma_k$ be the fraction of the parity-check equations of degree $k$, and $\Gamma(x) \triangleq \sum_k \Gamma_k x^k$ designate the corresponding right degree distribution of $\mathcal{C}$. Then the conditional entropy of the transmitted codeword given the received sequence satisfies

$$\frac{H(\mathbf{X}|\mathbf{Y})}{n} \geq 1 - C - (1-R)\left(1 - \frac{1}{2\ln(2)} \sum_{p=1}^{\infty} \frac{\Gamma(g_p)}{p(2p-1)}\right)$$

where

$$g_p \triangleq \int_0^{\infty} a(l)(1+e^{-l}) \tanh^{2p}\left(\frac{l}{2}\right) dl, \quad p \in \mathbb{N}, \quad (1)$$

and $a$ is the conditional pdf of the LLR given the channel input is 1 (where the possible inputs to the channel are $\{+1, -1\}$).

From Fano's inequality and since the code is binary, the conditional entropy of the transmitted codeword given the received sequence satisfies

$$\frac{H(\mathbf{X}|\mathbf{Y})}{n} \leq Rh(P_{\text{b}}) \qquad (2)$$

where $R$ is the rate of the code, $P_{\text{b}}$ is the bit-error probability of the ML decoder, and $h(x) = -x\log_2(x) - (1-x)\log_2(1-x)$ is the binary entropy function to base two. By combining Theorem 2.1 and Eq. (2), it is possible to derive an upper bound on the achievable rates of sequences of binary linear block codes and a lower bound on their bit-error probability, these bounds are presented in the next corollaries.

*Corollary 2.1:* Let $\{\mathcal{C}_m\}$ be a sequence of binary linear block codes whose codewords are transmitted with equal probability over an MBIOS channel, and assume that the block lengths of these codes tend to infinity as $m \to \infty$. Let $\Gamma_{k,m}$ be the fraction of the parity-check nodes of degree $k$ for an arbitrary representation of the code $\mathcal{C}_m$ by a bipartite graph which corresponds to a full-rank parity-check matrix, and assume the limit $\Gamma_k \triangleq \lim_{m\to\infty} \Gamma_{k,m}$ exits. Then in the limit where $m \to \infty$, a necessary condition on the asymptotic achievable rate $(R)$ for obtaining

vanishing bit error probability is

$$R \leq 1 - \frac{1-C}{1 - \frac{1}{2\ln(2)} \sum_{p=1}^{\infty} \frac{\Gamma(g_p)}{p(2p-1)}}.$$

*Corollary 2.2:* Let $\mathcal{C}$ be a binary linear block code of rate $R$ whose transmission takes place over an MBIOS channel with capacity $C$. For an arbitrary full-rank parity-check matrix $H$ of the code $\mathcal{C}$, let $\Gamma_k$ designate the fraction of parity-check equations that involve $k$ variables, and $\Gamma(x) \triangleq \sum_k \Gamma_k x^k$ be the right degree distribution from the node perspective which refers to the corresponding bipartite graph of $\mathcal{C}$. Then, under ML decoding (or any other decoding algorithm), the bit error probability ($P_\text{b}$) of the code satisfies

$$h_2(P_\text{b}) \geq 1 - \frac{C}{R} + \frac{1-R}{2R\ln(2)} \sum_{p=1}^{\infty} \frac{\Gamma(g_p)}{p(2p-1)}.$$

By combining Theorem 2.1 and Eq. (2), one can also derive a lower bound on the asymptotic parity-check density of a sequence of LDPC codes which achieve a fraction $1 - \varepsilon$ with vanishing bit-error probability. Similarly to the bound of Sason and Urbanke this bound is also of the form $\frac{\widetilde{K}_1 + \widetilde{K}_2 \ln \frac{1}{\varepsilon}}{1-\varepsilon}$, but the new bound is tighter then the one given in [21]. Particularly, the coefficient of the logarithmic growth rate $\widetilde{K}_2$ is strictly greater than the coefficient $K_2$ in the previous bound for all channels except the BSC (where the two-level quantization does not change the channel). It is also interesting to note that for the BEC the new bound merges with the improved bound of Sason and Urbanke which was shown to be tight even under MPI decoding.

The requirement of full-rank parity-check matrices is not natural when dealing with ensembles of codes; from [2], it can be removed by replacing the rate by the design rate of the ensemble.

**Numerical Results**: Table I provides bounds on the thresholds under ML decoding of several irregular rate $\frac{1}{2}$ ensembles taken from [20], where the transmission takes place over the binary input AWGN channel. The upper bounds on the achievable rates provide lower bounds on the $\frac{E_\text{b}}{N_0}$ thresholds under ML decoding. The bounds shown are the two-level quantization bound of Burshtein *et al.* [2], the four and eight-level quantization bounds based on the first approach discussed here, which are derived in [26, Corollaries 3.1, 3.2], and the un-quantized bound of Theorem 2.1. Comparing these bounds with the exact thresholds under MPI decoding calculated using density evolution (DE) shown in the rightmost column, gives an indication of the loss in performance due to the sub-optimality of the iterative decoder. As mentioned above, comparing the results of the quantized bounds with the un-quantized bound gives an indication on the effect of quantization at the decoder. Table I shows that the numerical values of the bounds which correspond to the optimized eight quantization levels are very close to the bounds where there is no quantization of the LLR.

## 3 Parallel Channels

In this section we consider LDPC codes transmitted over a set of independent parallel MBIOS channels, where each code bit is a-priori assigned to one of the channels. This setup is useful for modelling a wide range of scenarios which include non-uniform error protection [16], punctured LDPC codes (see e.g., [5], [17]) and LDPC-coded modulation schemes. Non-uniform error protection is modelled by assigning the bits which require a higher level of protection to a channel which is physically degraded w.r.t. the actual communication channel. Puncturing is modelled by assigning a code bit which is punctured in probability $p$ to the channel composed of a BEC with erasure probability $p$ cascaded with the actual communication channel. In [23] the authors generalized the lower bound in Theorem 2.1 for the case of parallel channels. It can also be verified that (2) holds for this case. Hence, the bounds on the achievable rates, the bit-error probability and the parity-check density discussed in the previous section are also generalized for the case of parallel channels. This allows us to study the performance of punctured LDPC codes transmitted over MBIOS channels.

## 4 Punctured LDPC codes

As discussed in the previous section, punctured codes transmitted over a single MBIOS channel can be modelled as codes transmitted over a set of parallel channels (see also [16]). The iterative decoder of the punctured code utilizes the graph of the original code, where the nodes corresponding to the punctured code bits receive no channel input and are considered state nodes. This allows to use puncturing in order to transmit at variable rates using one encoder and one decoder. Some fundamental properties of punctured LDPC codes were studied in [17]. These results show the high potential of puncturing in designing rate-compatible codes which operate closely to the channel capacity.

### 4.1 Intentionally Punctured LDPC codes

In [5], Ha and McLaughlin show that good codes can be constructed by puncturing good ensembles of LDPC codes using a technique called "intentional puncturing". In this approach, the code bits are partitioned into disjoint sets so that each set contains all the code bits whose corresponding variable nodes have the same degree. The code bits in each one of these sets are randomly punctured at a fixed puncturing rate. The puncturing pattern is described by a polynomial $\pi_0(x) = \sum_{i \geq 2} \pi_i x^{i-1}$ where $\pi_i$ designates the puncturing rate of code bits of degree $i$. This forms a particular case of transmission over a set of parallel channels, where each code bit is assigned to a channel according to the number of parity-check equations it is involved in. The bound on the achievable rates of ML decoded

| $\lambda(x)$ | $\rho(x)$ | 2-Level Bound [2] | 4-Level Bound | 8-Level Bound | Un-Quantized Lower Bound | DE Threshold |
|---|---|---|---|---|---|---|
| $0.38354x + 0.04237x^2 + 0.57409x^3$ | $0.24123x^4 + 0.75877x^5$ | 0.269 dB | 0.370 dB | 0.404 dB | 0.417 dB | 0.809 dB |
| $0.23802x + 0.20997x^2 + 0.03492x^3 + 0.12015x^4 + 0.01587x^6 + 0.00480x^{13} + 0.37627x^{14}$ | $0.98013x^7 + 0.01987x^8$ | 0.201 dB | 0.226 dB | 0.236 dB | 0.239 dB | 0.335 dB |
| $0.21991x + 0.23328x^2 + 0.02058x^3 + 0.08543x^5 + 0.06540x^6 + 0.04767x^7 + 0.01912x^8 + 0.08064x^{18} + 0.22798x^{19}$ | $0.64854x^7 + 0.34747x^8 + 0.00399x^9$ | 0.198 dB | 0.221 dB | 0.229 dB | 0.232 dB | 0.310 dB |
| $0.19606x + 0.24039x^2 + 0.00228x^5 + 0.05516x^6 + 0.16602x^7 + 0.04088x^8 + 0.01064x^9 + 0.00221x^{27} + 0.28636x^{29}$ | $0.00749x^7 + 0.99101x^8 + 0.00150x^9$ | 0.194 dB | 0.208 dB | 0.214 dB | 0.216 dB | 0.274 dB |

TABLE I

COMPARISON OF THRESHOLDS FOR RATE ONE-HALF ENSEMBLES OF IRREGULAR LDPC CODES TRANSMITTED OVER THE BINARY-INPUT AWGN CHANNEL. THE SHANNON CAPACITY LIMIT CORRESPONDS TO $\frac{E_b}{N_0} = 0.187$dB. THE 2-LEVEL, 4-LEVEL, 8-LEVEL AND UN-QUANTIZED LOWER BOUNDS ON THE THRESHOLD REFER TO ML DECODING, AND ARE BASED ON [2, THEOREM 2], [26, COROLLARIES 3.1, 3.2] AND THEOREM 2.1, RESPECTIVELY. THE DEGREE DISTRIBUTIONS OF THE ENSEMBLES AND THEIR DE THRESHOLDS ARE BASED ON DENSITY EVOLUTION UNDER SUM-PRODUCT DECODING [19], AND ARE TAKEN FROM [20, TABLES 1 AND 2].

| $\pi^0(x)$ (puncturing pattern) | Design rate | Capacity limit | Lower bound (ML decoding) | Iterative (IT) Decoding | Fractional gap to capacity (ML vs. IT) |
|---|---|---|---|---|---|
| 0 | 0.500 | 0.187 dB | 0.270 dB | 0.393 dB | $\geq 40.3\%$ |
| $0.07886x + 0.01405x^2 + 0.06081x^3 + 0.07206x^9$ | 0.528 | 0.318 dB | 0.397 dB | 0.526 dB | $\geq 37.9\%$ |
| $0.20276x + 0.09305x^2 + 0.03356x^3 + 0.16504x^9$ | 0.592 | 0.635 dB | 0.716 dB | 0.857 dB | $\geq 36.4\%$ |
| $0.25381x + 0.15000x^2 + 0.34406x^3 + 0.019149x^9$ | 0.629 | 0.836 dB | 0.923 dB | 1.068 dB | $\geq 37.3\%$ |
| $0.31767x + 0.18079x^2 + 0.05265x^3 + 0.24692x^9$ | 0.671 | 1.083 dB | 1.171 dB | 1.330 dB | $\geq 35.6\%$ |
| $0.36624x + 0.24119x^2 + 0.49649x^3 + 0.27318x^9$ | 0.719 | 1.398 dB | 1.496 dB | 1.664 dB | $\geq 36.9\%$ |
| $0.41838x + 0.29462x^2 + 0.05265x^3 + 0.30975x^9$ | 0.774 | 1.814 dB | 1.927 dB | 2.115 dB | $\geq 37.2\%$ |
| $0.47074x + 0.34447x^2 + 0.02227x^3 + 0.34997x^9$ | 0.838 | 2.409 dB | 2.547 dB | 2.781 dB | $\geq 37.1\%$ |
| $0.52325x + 0.39074x^2 + 0.01324x^3 + 0.39436x^9$ | 0.912 | 3.399 dB | 3.607 dB | 3.992 dB | $\geq 35.1\%$ |

TABLE II

COMPARISON OF THRESHOLDS FOR ENSEMBLES OF INTENTIONALLY PUNCTURED LDPC CODES WHERE THE ORIGINAL ENSEMBLE BEFORE PUNCTURING HAS THE DEGREE DISTRIBUTIONS $\lambda(x) = 0.25105x + 0.30938x^2 + 0.00104x^3 + 0.43853x^9$ AND $\rho(x) = 0.63676x^6 + 0.36324x^7$ (SO ITS DESIGN RATE IS EQUAL TO $\frac{1}{2}$). THE TRANSMISSION TAKES PLACE OVER A BINARY-INPUT AWGN CHANNEL. THE TABLE COMPARES VALUES OF $\frac{E_b}{N_0}$ REFERRING TO THE CAPACITY LIMIT, THE GENERALIZATION TO PARALLEL CHANNEL OF THE BOUND GIVEN IN COROLLARY 2.1 (WHICH PROVIDES A LOWER BOUND ON $\frac{E_b}{N_0}$ UNDER ML DECODING), AND THRESHOLDS UNDER ITERATIVE MESSAGE-PASSING DECODING. THE FRACTIONAL GAP TO CAPACITY (SEE THE RIGHTMOST COLUMN) MEASURES THE RATIO OF THE GAP TO CAPACITY UNDER OPTIMAL ML DECODING AND THE ACHIEVABLE GAP TO CAPACITY UNDER (SUB-OPTIMAL) ITERATIVE MESSAGE-PASSING DECODING. THE DEGREE DISTRIBUTIONS FOR THE ENSEMBLE OF LDPC CODES, AND THE POLYNOMIALS WHICH CORRESPOND TO ITS PUNCTURING PATTERNS ARE TAKEN FROM [5, TABLE 2].

LDPC codes transmitted over parallel channels, can be used to find lower bounds on the thresholds which intentionally punctured LDPC codes achieve under ML decoding. Comparing these bounds with exact thresholds calculated in [5] using DE, gives an indication as to the inherent loss caused by the sub-optimality of the iterative decoder for various puncturing patterns and rates. The bound also gives an indication of the loss of punctured codes even under ML decoding due to the structure of the codes (as compared to random coding). Table II presents the lower bound on the threshold under ML decoding of a sequence of ensembles of LDPC codes with various puncturing rates, where the transmission takes place over the binary input AWGN channel. It can be seen that for all the puncturing rates shown in Table II, more than $\frac{1}{3}$ of the gap to capacity is attributed to the structure of the codes, while the rest is due to the sub-optimality of the iterative decoder.

### 4.2 Randomly punctured LDPC codes

A second approach to the construction of ensembles of punctured LDPC codes is to a-priori select a subset of the code bits and randomly puncture these bits at a given rate. This method can also be modelled as a specific case of transmission over a set of two parallel channels. By observing the lower bound on the parity-check density for parallel channels, one is able to derive a lower bound on the decoding complexity per iteration of randomly punctured LDPC codes.

In [14], Pfister et al. provided explicit constructions of ensembles of irregular repeat-accumulate (IRA) codes which asymptotically (as the block length tends to infinity) achieve the capacity of the BEC under MPI decoding with bounded encoding and decoding complexities per information bit. This result came as a surprise in light of parallel results which show that for ensembles of LDPC codes without puncturing and *systematic* IRA codes, their decoding complexity per information bit should become unbounded as the gap to capacity vanishes (see [21] and [22]). Since the convergence rate (in terms of the block length) of the codes in [14] was rather slow, these codes were not record breaking for short to moderate block lengths. Several constructions of ensembles of systematic accumulate-repeat-accumulate (ARA) codes were recently introduced by Pfister and Sason [15]. These ensembles achieve capacity of the BEC under MPI decoding with bounded complexity per information bit, and outperform other capacity-achieving ensembles of codes on graphs, even for moderate block lengths. In [7], Hsu and Anastasopoulos introduced ensembles of codes which achieve the capacity of general MBIOS channels under ML decoding with bounded number of edges in the Tanner graph per information bit, and for the BEC, they also achieve bounded complexity per information bit under MPI decoding. All of the capacity-achieving ensembles on the BEC with bounded complexity rely on puncturing, as could be expected from the information-theoretic bounds for codes on graphs in [13], [14, Theorems 3 and 4] and [23].

**Acknowledgment**: The work was supported by the EU 6$^{\text{th}}$ International Framework Programme via the NEWCOM Network of Excellence.